\documentclass[conference,compsoc]{IEEEtran}
\usepackage{xspace}
\usepackage{tikz}
\usepackage{booktabs}
\usepackage{amssymb}

\usepackage{amsmath}
\usepackage{subcaption}
\usepackage{stfloats}
\usepackage{xcolor, soul}
\usepackage{xurl}
\usepackage[ruled,vlined,linesnumbered]{algorithm2e}
\usepackage{lipsum}
\usetikzlibrary{positioning,arrows.meta}
\AtBeginDocument{}

\usepackage{pgfplots}
\pgfplotsset{compat=1.18}

\newcommand{\acro}{\textsc{VeCoDI}\xspace}
\newcommand{\iobox}{\textsc{Shangri-La}\xspace}

\newcommand{\dev}{\textsc{Dev}\xspace}
\newcommand{\modOwner}{\textsc{Pvd}\xspace}
\newcommand{\modUser}{\textsc{Csm}\xspace}
\newcommand{\modID}{\textsc{$ID_M$}\xspace}

\newcommand{\vrf}{\textsc{Vrf}\xspace}
\newcommand{\f}{\textsc{F}\xspace}
\newcommand{\m}{\textsc{$M$}\xspace}
\newcommand{\mPriv}{\textsc{$M_{pri}$}\xspace}
\newcommand{\mPub}{\textsc{$M_{pub}$}\xspace}
\newcommand{\authToken}{\textsc{$T_{\modUser}$}\xspace}
\newcommand{\infToken}{\textsc{$T_{proof}$}\xspace}
\newcommand{\fIn}{\textsc{$In_\f$}\xspace}
\newcommand{\fOut}{\textsc{$Out_\f$}\xspace}

\newcommand{\advMod}{\ensuremath{\mathcal{A_{M}}}\xspace}
\newcommand{\advInf}{\ensuremath{\mathcal{A_{I}}}\xspace}

\newcommand\ignore[1]{}

\newcommand{\pmark}{$\sim$} 
\sethlcolor{grey}

\title{\acro: Verifiable and Confidential DNN Inference on Low-End Edge Devices}

\author{\IEEEauthorblockN{Mohamed Khalil Kiri\IEEEauthorrefmark{1},
Ivan De Oliveira Nunes\IEEEauthorrefmark{2},
Aurélien Francillon\IEEEauthorrefmark{1}, and 
Norrathep Rattanavipanon\IEEEauthorrefmark{3}}
\IEEEauthorblockA{\IEEEauthorrefmark{1}EURECOM, France\\
\IEEEauthorrefmark{2}University of Zurich, Switzerland\\
\IEEEauthorrefmark{3}Prince of Songkla University (PSU), Phuket, Thailand}
}

\begin{document}

\maketitle
\begin{abstract}
Deploying deep neural network (DNN) inference on low-end edge devices raises two key challenges: protecting model confidentiality against a potentially compromised edge system and enabling verifiable inference without incurring prohibitive overhead.
Existing approaches either house (partial) models and inference software within trusted execution environments (TEEs), resulting in high cost and an application-dependent trusted computing base (TCB), or execute in untrusted environments, providing little security.

In this work, we present \acro, a framework for verifiable and confidential DNN inference on constrained edge devices.
At its core, \acro introduces \iobox, a new execution abstraction on TrustZone-M TEEs that establishes a third runtime environment with privileges strictly between the Secure and Non-Secure Worlds.
\acro leverages \iobox to execute untrusted inference code in the Non-Secure World while using minimal application-agnostic Secure-World support to protect model confidentiality and enable verifiability (with respect to proper execution of inference code and model parameters) of inference results.
We realize \acro on a real-world NUCLEO-L552ZE-Q development board and open-source its prototype.
Our results show \acro's small TCB, memory footprint, and runtime overhead, making it a practical option for secure inference in low-end edge devices.
\end{abstract}




\section{Introduction}

Edge computing has emerged as a paradigm for performing inference with machine learning (ML) models, particularly deep neural networks (DNNs), in latency-sensitive and resource-constrained environments~\cite{li2019edge}. 
In this paradigm, DNN models are trained in centralized infrastructures (e.g., the cloud) by service providers, referred to as model providers, and subsequently deployed to edge devices, such as sensor and actuator microcontrollers.
These low-cost devices then perform DNN inference locally, enabling real-time decision-making for several modern IoT applications such as remote medical diagnostics, autonomous systems, and industrial monitoring~\cite{li2019edge}.
However, moving from centralized to decentralized deployments creates new attack vectors.

First, training DNN models typically requires substantial computational resources, as well as significant effort in collecting and curating training data. 
As such, model providers often consider these models to be valuable intellectual property/assets that must be protected~\cite{xue2021intellectual}. 
However, when deployed at the edge, DNN models are exposed to stronger adversarial capabilities, where device-controlled adversaries have both physical and software access to the device. 
This enables such adversaries to: (1) directly extract the model, either at rest (e.g., from on-device flash memory) or at runtime, and (2) mount \emph{model extraction} attacks~\cite{tramer2016stealing} by reconstructing the deployed model via performing a large number of carefully crafted inference queries. 
These threats introduce a new challenge of \emph{model confidentiality} for model providers in edge deployments.

Second, edge inference outcomes are often used to drive critical decision-making. 
For example, in autonomous systems, object detection results are used to identify obstacles (e.g., pedestrians) and trigger braking upon detection. 
Similarly, in remote medical diagnostics, ECG analysis results may be used to detect cardiac abnormalities and initiate timely intervention~\cite{li2019edge}. 
However, edge devices operate under strict constraints in terms of cost, size, and power, which limit the deployment of sophisticated security mechanisms. 
As a result, these low-end devices are more vulnerable to malware infestation, leading to a surge in IoT malware targeting edge devices in recent years~\cite{pinto2019demystifying}. 
Once compromised, such malware can take control of the entire software stack and produce spoofed inference results to manipulate downstream decisions, potentially endangering physical environments and users. 
To address this second class of attacks, \emph{verifiable inference} is necessary. In traditional centralized settings, this is typically not a major concern, as the model providers have full control over powerful inference devices.

To enable \emph{model confidentiality} with low performance overhead, recent work leverages Trusted Execution Environments (TEEs), giving rise to a class of techniques known as \emph{TEE-Shielded DNN Partitioning (TSDP)}~\cite{li2025teeslice,hou2021model,mo2020darknetz,sun2023shadownet}. 
These techniques partition a DNN into two disjoint components: (i) \emph{privacy-sensitive}
and (ii) \emph{privacy-insensitive} parts.
The privacy-sensitive components are executed within the TEE, while the remaining parts are left unprotected. 
Despite good performance, this line of work suffers from a key limitation: it enlarges the trusted computing base (TCB) by incorporating DNN inference code directly into the TEE, thereby increasing both the software footprint and the attack surface to already-constrained/vulnerable edge devices.
Moreover, many of these approaches~\cite{hou2021model,mo2020darknetz,sun2023shadownet} overlook model extraction attacks, leaving this important issue unaddressed.

On the other hand, prior work on \emph{verifiable inference} either assumes that the entire inference pipeline is executed within TEEs~\cite{volos2018graviton,duddu2024laminator,chantasantitam2026pal} or relies on cryptographic techniques in conjunction with TEEs~\cite{tramer2018slalom,zhang2024verisplit}. 
The former is often undesirable in edge settings, as DNN models may exceed the limited memory capacity of the TEE.
The latter, while more flexible, can introduce significant computational overhead, making it unsuitable for power-constrained devices. 
More importantly, both approaches still suffer from TCB bloat due to the inclusion of complex and potentially buggy inference code within the TEE. 
A compromise of such code could escalate into other high-assurance components (e.g., secure key storage, secure software update) that live in the same TEE, as is the case with ARM TrustZone-M -- a commonly used TEE for low-end edge devices.



In light of these challenges, we introduce \acro, a TrustZone-M-based framework that 
simultaneously enables \emph{model confidentiality} and \emph{verifiable inference} for DNNs on low-end edge devices. 
%
%
Unlike prior TEE-based approaches, \acro does not place DNN inference code or data in the Secure World. Instead, it introduces a new abstraction, \iobox, which serves as a third runtime environment alongside TrustZone-M's Secure and Non-Secure Worlds.

\iobox resides in the Normal World memory but operates with elevated privileges enforced by the Secure World. This design enables reuse of existing Normal World functionality, avoiding cross-world code duplication. At the same time, \iobox guarantees the confidentiality and integrity of its code and data -- both at rest and during execution -- even when the Normal World is untrusted.
In addition, \iobox can directly access hardware I/O peripherals, preventing interference from the Normal World on the physical acquisition of data. Despite these capabilities, it remains less privileged than the Secure World, ensuring that a compromised or faulty \iobox cannot impact Secure World components.
Finally, \iobox requires only minimal, application-agnostic code in the Secure World. This code can be thoroughly tested or formally verified once and reused across multiple \iobox applications.

At the protocol level, we combine \iobox isolation abstraction with two key features. 
First, an inference authorization protocol allows the model provider to limit the number/frequency of inferences that can be performed;
this mitigates \emph{model extraction attacks} that threaten \emph{model confidentiality}. 
Second, we extend \iobox to generate a verifiable proof of inference execution that accompanies each inference result. 
This proof allows an external verifier to assess whether: (1) the inference inputs were correctly captured from physical interfaces; (2) inference code is executed correctly on the intended DNN model and authentic input; and (3) the reported inference result corresponds exactly to this execution instance.
This also supports privacy, since the proof validation does not require raw inputs used by the edge device that may contain sensitive user information (e.g., consider an image recognition case).

\textbf{\bf Contributions}
Our main contributions are as follows:

    


    \noindent \textbf{1- Third Runtime Environment.}
    We introduce \iobox, a lightweight execution abstraction on top of TrustZone-M.
    It provides a third runtime environment that is protected from Normal World compromise, while ensuring that any compromise within \iobox cannot affect the Secure World. 
    At the same time, \iobox physically lives in Normal World and thus can reuse existing Normal World code, reducing the overall software footprint.  
    We believe \iobox is of independent interest beyond secure DNN inference, as it enables third-party developers to deploy arbitrary trusted applications on TrustZone-M without modifying core Secure World components.

    \noindent \textbf{2- Secure and Verifiable DNN Inference.}
    We build upon the \iobox abstraction to achieve both \emph{model confidentiality} and \emph{verifiable inference} on low-end edge devices.
    First, we equip \iobox with an inference authorization mechanism to mitigate model extraction attacks.
    Second, we enable \iobox to generate verifiable proofs of input retrieval and inference execution, allowing external verifiers to validate results without accessing sensitive raw input data.
    Along with these extensions, \iobox is used to construct \acro: a protocol executed between inference stakeholders to realize confidential and verifiable DNN inference.

    \noindent \textbf{3- Practical Deployment and Evaluation.}
    We implement \acro on a NUCLEO-L552ZE-Q development board~\cite{st_nucleo_l552ze_q} and open-source its prototype~\cite{anon_repo}. 
    Our evaluation results demonstrate the practical benefits of \acro: 
    Compared to prior TEE-based solutions, \acro eliminates the need to place inference code in the Secure World, reducing the TCB by 95.64\%. 
    Moreover, \acro incurs only 0.83 ms (0.07\%) runtime overhead, making it well-suited for resource-constrained devices. 
    Finally, we demonstrate real-world applicability through a case study on camera-based edge image classification.


\section{Background}\label{sec:bg}


\subsection{TrustZone for ARM Cortex-M MCUs}

ARM TrustZone-M provides hardware-enforced isolation for Cortex-M microcontrollers by partitioning the system into two domains: the Secure World and the Normal World.
Each domain can only access memory regions assigned to its corresponding security state.

Memory regions, including code, data, and peripherals, are classified as Secure, Non-Secure, or Non-Secure Callable (NSC) through the Security Attribution Unit (SAU) and the Implementation-Defined Attribution Unit (IDAU), both configured by Secure-World code at boot time.
This ensures that only Secure-World code can access Secure memory, while Non-Secure code is restricted to Non-Secure regions.

To enable controlled invocation, the Secure World exposes APIs that can be called by the Normal World through designated NSC entry points.
These entry points are the only locations where control can transfer from the Non-Secure World to the Secure World, preventing arbitrary jumps into Secure-World code.
As a result, the Normal World cannot tamper with the execution of Secure-World code except through explicitly defined APIs, and cannot access Secure-World resources protected by the SAU and IDAU, ensuring strong isolation between these two worlds.

In practice, the Secure World contains a \emph{small} trusted computing base (TCB) consisting of security-critical components (e.g., secure boot, key storage, and attestation) to minimize the attack surface.
In contrast, the Non-Secure World hosts a significantly larger software stack, including the operating system and user applications.

We also note that, unlike the IDAU, the SAU can be reconfigured at runtime by the Secure World, allowing memory regions to transition dynamically between Secure and Non-Secure states.


\subsection{Remote Attestation and Proof of Execution}

\textbf{Remote Attestation (RA).} RA allows a remote verifier to establish trust in the
software state of a target device~\cite{ menetrey2022attestation}. 
To be secure, RA requires a root of trust (RoT) on the target device (e.g., TrustZone-M TEE) to measure the software to be attested and sign (or MAC) the measurement result under a device-specific attestation key, yielding attestation evidence.
The verifier can validate the evidence by checking the signature/MAC using the pre-shared verification key and matching the measurement against a known-good (golden) reference.
While RA certifies \emph{what} software is installed, it offers no guarantee over \emph{what the device subsequently computes}.

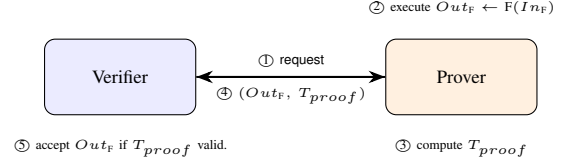
\begin{figure}[t]
    \centering
    \begin{tikzpicture}[
        node distance=10mm and 20mm,
        entity/.style={draw, rounded corners, align=center,
                       minimum width=20mm, minimum height=10mm,
                       font=\scriptsize, fill=blue!8},
        prover/.style={draw, rounded corners, align=center,
                       minimum width=20mm, minimum height=10mm,
                       font=\scriptsize, fill=orange!12},
        arr/.style={-{Stealth[length=2.2mm]}, line width=0.75pt},
        lbl/.style={font=\tiny}
    ]
        \node[entity] (vrf) {Verifier};
        \node[prover, right=25mm of vrf] (prv) {Prover};

        \draw[arr] (vrf.east) -- node[above,lbl]{%
            \textcircled{\tiny 1}~\textsf{request}} (prv.west);

        \node[lbl, above=2mm of prv, align=center]
        {\textcircled{\tiny 2}~execute $Out_\f \leftarrow \f(In_\f)$};

        \node[lbl, below=2mm of prv, align=center]
        {\textcircled{\tiny 3}~compute \infToken};

        \draw[arr] (prv.west) -- node[below,lbl]{%
            \textcircled{\tiny 4}~$(\fOut,\,\infToken)$} (vrf.east);

        \node[lbl, below=2mm of vrf, align=center]
        {\textcircled{\tiny 5}~accept $Out_\f$ if $\infToken$ valid.};
    \end{tikzpicture}
    \caption{Generic Proof-of-Execution (PoX) workflow.}
    \label{fig:pox_workflow}
\end{figure}

\textbf{Proof of Execution (PoX).} PoX extends RA evidence to include execution information~\cite{nunes2020apex}.
The RoT attests not only to code identity but to the fact that a designated function
\f ran to completion on a given input and produced a given output within the
attested device. As illustrated in Figure~\ref{fig:pox_workflow}, the protocol
follows a challenge-response exchange between a verifier and a prover device:

    \noindent \textbf{1- Request.} Verifier issues \textsf{request} containing a cryptographic challenge. It also specifies which function \f should be executed and optionally a verifier-provided input $In_\f$.

    \noindent \textbf{2- Execution.} Upon receiving it, the prover computes $Out_\f \leftarrow \mathsf{\f}(In_\f)$; the RoT monitors this execution and immediately aborts it if the execution is tampered with. Note that \f code may contain direct access to hardware I/O peripherals for data acquisition (i.e., it should contain it when performing inference on an edge sensor device).

    \noindent \textbf{3- Proof generation.} The RoT produces attestation evidence \infToken. This token should be bound to the challenge received in the request, the executed \f binary, its legal entry point, $In_\f$, and $Out_\f$. Crucially, \infToken should only be produced if \f execution is completed successfully and (for timeliness) after the receipt of the challenge it is bound to.

    \noindent \textbf{4- Response.} The prover returns $(\fOut,\,\infToken)$;
    The verifier accepts $Out_\f$ as authentic if and only if \infToken is successfully verified using the verification key and the measurement matches the expected challenge, \f binary, received $Out_\f$, and expected input $In_\f$ (if any).

\subsection{TEE-Shielded DNN Partitioning}

Running a full DNN inside a low-end TrustZone-M TEE is generally infeasible: Secure RAM is measured in tens of kilobytes, whereas even compact models require orders of magnitude more memory. 
TEE-Shielded DNN Partitioning (TSDP) addresses this by
decomposing a model $M$ into two sub-networks: 
a sensitive component $M_\mathit{pri}$ sequestered in the Secure
World, and a public component $M_\mathit{pub}$ executed in the Normal World under
the device's full memory and compute budget. 
The choice of boundary is dictated by the target security property, e.g., model protection~\cite{li2025teeslice,10247898} or execution efficiency~\cite{10949698}.

\ignore{

\textbf{Partition objectives.} 
The choice of boundary is dictated by the target security property:

    \noindent \textbf{- IP protection.} TEESlice~\cite{li2025teeslice} and
    SecureQNN~\cite{11130497} isolate bottleneck layers whose parameters
    are critical to model utility, raising the cost of surrogate-model reconstruction
    via extraction or distillation attacks~\cite{10247898}.

    \noindent \textbf{- Input privacy.} ShadowNet~\cite{sun2023shadownet} and
    SelTZ~\cite{jeong2024seltz} shield the initial layers, which are most susceptible to
    feature-inversion and membership-inference attacks against raw input data.

    \noindent \textbf{- Execution efficiency.} SmartZone~\cite{10949698} minimizes
    world-switch frequency by colocating layers that exchange large activation tensors,
    bounding latency to the number of partition crossings rather than model depth.
}

Regardless of objective, TSDP requires hosting $M_\mathit{pri}$ alongside
a full inference engine (e.g., TFLite Micro) inside the Secure World, inflating the
TCB beyond the minimality required for a secure RoT. End-to-end latency is similarly
constrained: on cycle-limited MCUs, fine-grained partitions that maximize security
coverage do so at the cost of frequent pipeline stalls on each domain crossing.

\section{System/Threat Models and Goals}\label{sec:prelim}

\ignore{
\subsection{Motivating Examples}

We present two application use cases that motivate the need for \emph{model confidentiality} and \emph{verifiable inference} for deploying DNN inference in edge settings.

First, in medical applications, diagnostic models are increasingly deployed on portable devices such as vital signal monitors (e.g., ECG or heartbeat)~\cite{} and imaging systems~\cite{}. 
These models embody significant intellectual property, making them attractive targets for adversaries who may purchase such devices to extract the models and use them for their benefits. 
At the same time, the authenticity of inference results is crucial for healthcare providers to perform remote diagnosis correctly. 
They therefore require strong guarantees that inference results are generated by an authentic, approved model and based on untampered inference input.
Also, since the inference input -- the user's health data -- is highly sensitive, these guarantees must be provided without revealing it.

Second, in autonomous driving and advanced driver-assistance systems (ADAS), perception models deployed on vehicles make real-time decisions with direct safety implications. 
These models are commercially sensitive. 
Without model confidentiality, an adversary may extract these valuable models for competitive use. 
Without verifiable inference, an adversary may forge inference results sent to the decision-making ECU, potentially bypassing safety mechanisms and causing physical harm to the vehicle, its occupants, or the surrounding environment.
}

\subsection{System and Threat Models}\label{subsec:system-threat-model}

\textbf{System Model.} We consider an edge inference setting that comprises four stakeholders:

\noindent \textbf{- Model provider} (\modOwner): curates training data, provides the DNN model \m, and authorizes access to it.

\noindent \textbf{- Edge device} (\dev): a single-core MCU equipped with a low-cost TrustZone-M TEE. Note that MCUs lack memory management units, making them unsuitable for isolation mechanisms based on virtual memory (e.g., operating system- or hypervisor-based). Our model reflects this.

\noindent \textbf{- Model consumer} (\modUser): obtains authorization from \modOwner to run inference on \dev.

    
\noindent \textbf{- Verifier} (\vrf): receives inference results and validates their trustworthiness before downstream use. Although logically distinct, \vrf's role can be implemented by another entity, e.g., in practice, a single entity can play both \vrf and \modUser roles, where the consumer of a result also locally verifies it.

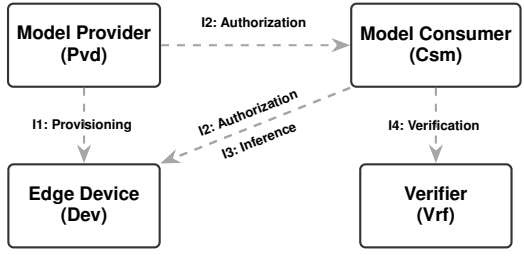
\begin{figure}[t]
    \centering
    \begin{tikzpicture}[
        node distance=1.5cm and 2.5cm,
        entity/.style={draw=black!80, fill=white, thick, rounded corners=2pt, align=center, minimum width=2cm, minimum height=1.1cm, font=\scriptsize\sffamily\bfseries},
        setup_flow/.style={-{Stealth[length=2.5mm]}, thick, dashed, draw=gray!70},
        exec_flow/.style={-{Stealth[length=2.5mm]}, thick, draw=black!90},
        lbl/.style={font=\tiny\sffamily, align=center}
    ]

    \node[entity] (pvd) {Model Provider \\ (\textbf{Pvd})};
    \node[entity, right=2.5cm of pvd] (csm) {Model Consumer \\ (\textbf{Csm})};
    
    \node[entity, below=1cm of pvd] (dev) {Edge Device \\ (\textbf{Dev})};
    \node[entity, below=1cm of csm] (vrf) {Verifier \\ (\textbf{Vrf})};

    \draw[setup_flow] (pvd.south) -- node[lbl, midway] {
        \textbf{I1: Provisioning} 
    } (dev.north);

    \draw[setup_flow] (pvd.east) -- node[lbl, midway, above=3pt] {
        \textbf{I2: Authorization} 
    } (csm.west);

    
    \draw[setup_flow] (csm.south west) -- node[lbl, midway, sloped, below] {
        \textbf{I3: Inference} 
    } (dev.north east);
    \draw[setup_flow] (csm.south west) -- node[lbl, midway, sloped, above] {
        \textbf{I2: Authorization} 
    } (dev.north east);
    

    \draw[setup_flow] (csm.south) -- node[lbl, midway] {
        \textbf{I4: Verification} 
    } (vrf.north);


    \end{tikzpicture}
    \caption{Interactions between \modOwner, \dev, \modUser, and \vrf.}
    \label{fig:protocol-overview}
\end{figure}
As illustrated in Figure~\ref{fig:protocol-overview}, secure edge inference consists of four interactions among these stakeholders:

 \noindent  \textbf{I1: Provisioning.} \modOwner securely provisions \dev with required cryptographic material and installs the input-retrieval and inference code, \f, along with \m on \dev.

\noindent  \textbf{I2: Authorization.} \modUser requests permission from \modOwner to use \m on \dev. 
    If approved, \modOwner issues an authorization token \authToken on $M$'s identifier \modID, encoding \modUser's usage policy on $M$.

\noindent \textbf{I3: Inference.} \modUser invokes inference on \dev along with the input $In_\f$ and $T_{u}$ which is an authenticated token issued by \modUser to execute one inference instance. 
    Upon validation, \dev executes $\f(In_\f)$ with \m, producing an inference result $Out_\f$ along with a proof \infToken.

\noindent \textbf{I4: Proof Verification.} \modUser sends $Out_\f$ and \infToken to \vrf, which validates the authenticity of $Out_\f$ w.r.t. the expected $\f$ and \infToken, and accepts $Out_\f$ if verification succeeds.


\textbf{Threat Model.} We consider two adversary cases \advMod and \advInf with different goals.

The goal of \advMod is to extract \m. 
\advMod can compromise \dev (except for its TEE-protected code and data), \modUser, and \vrf. \advMod cannot compromise \modOwner who already possesses $\m$.
By compromising \dev, \advMod gains full control over its communication channels, enabling message interception, modification, replay, and injection. It also controls all untrusted software on \dev, with the ability to read, write, and execute arbitrary code and data in the Normal World. 
\advMod also has physical access to \dev, allowing it to extract sensitive information (e.g., a DNN model) that is stored unprotected in \dev's Flash memory (e.g., via Flash dump).

For \advInf, the goal is to convince \vrf to accept a forged inference result $Out_\f$, that is: one that was not produced by the timely execution of the correct inference software on \dev using appropriate model parameters and external inputs (if any). 
\advInf can compromise \dev (except for TEE-protected state), and \modUser to achieve its goal.

\textbf{Assumptions.}
We assume \modOwner is trusted: as it already possesses $\m$, it has no incentive to act as \advMod. Conversely, being the model provider, it can already choose \m, gaining no advantage from behaving as \advInf.


\dev is assumed to be equipped with a TrustZone-M TEE, whose Secure World code is trusted and free of vulnerabilities.
This assumption is corroborated by our design principle of avoiding a potentially large inference function $\f$ in the Secure World.
Following the typical TrustZone-M trust model~\cite{pinto2019demystifying}, we assume an underlying secure boot process that guarantees the integrity of Secure World code at boot time, as well as secure persistent storage restricted to Secure World accesses~\cite{gonzalez2014tee}.
To protect against attacks based on Direct Memory Access (DMA) controllers, we assume that \dev is equipped with a security DMA controller capable of specifying DMA access permissions for memory regions, such as the Global TrustZone Controller (GTZC) available on our NUCLEO-L552ZE-Q prototype board.

The trusted \modOwner is responsible for ensuring \f correctness, i.e., it correctly retrieves input data from properly configured peripherals and performs inference without leaking sensitive information (i.e., $M$). 
However, given its size, there is still a chance for \modOwner to make mistakes and introduce vulnerabilities in \f. 
In such a case, while it is not possible to guarantee inference integrity, \acro ensures that the Secure World remains unaffected (in contrast to the alternative of housing \f in the Secure World).

\textbf{Out-of-Scope Attacks.}
Following the standard TrustZone-M threat model~\cite{tfm_physical_attack},  we do not consider physical attacks such as side-channel attacks, fault injection, invasive hardware modification, or denial-of-service attacks. 
While powerful against this class of devices, these attacks are orthogonal to this work and require dedicated/independent countermeasures~\cite{tfm_physical_attack,pouyanrad2024automated,sass2023oops}.

\subsection{Goals}\label{sec:goals}

Under the above system and threat models, \acro aims to achieve:

\subsubsection*{\textbf{G1 (Model Confidentiality)}}
\advMod should not learn any information about $\m$ beyond what \modOwner allows. 
This goal consists of two sub-goals:

    \textbf{G1-1 (Direct Leakage Protection):} Even with compromised \dev, \advMod cannot access $\m$ or any function of $\m$, except for $Out_\f$ and \infToken.

    \textbf{G1-2 (Indirect Leakage Mitigation):} Since $Out_\f$ may still leak information about $\m$ (e.g., via model extraction attacks), \advMod must not be able to invoke \f more than what \modOwner explicitly authorizes.
    
 \textbf{Remark.} \acro deliberately leaves the number of authorized invocations configurable by \modOwner. This is appropriate because \modOwner understands the model and is best positioned to set this limit, which may vary across different $\m$-s and application scenarios. Choosing suitable limits is orthogonal to \acro's architectural support.

\subsubsection*{\textbf{G2 (Verifiable Inference)}}
An adversary \advInf must not be able to trick \vrf into accepting invalid inference results. 
It consists of two sub-goals:

    \textbf{G2-1 (Authenticity):} \vrf must be able to detect invalid pairs $(\fOut, \infToken)$ submitted by \advInf. 
    In particular, this happens when: (1) $Out_\f$ is forged without executing $\f$, (2) $Out_\f$ is produced by executing $\f$ on a different \dev, (3) $Out_\f$ is produced by executing \f on the right \dev but using a different model than \m, or (4) $Out_\f$ results from an incomplete execution of $\f$ on \dev or from the execution of code different from \f.

    \textbf{G2-2 (Privacy):} To preserve \dev owner's privacy, \vrf must be able to validate $(Out_\f, \infToken)$ without access to the inference input, which may contain sensitive data.

\noindent\textbf{Remark.} The inference input is distinct from $\fIn$.
The latter denotes a public input to $\f$, containing only non-sensitive information (e.g., preprocessing parameters or peripheral configuration settings), hence not critical to the \dev owner's privacy.
In contrast, the inference input is not a part of $\fIn$; it is acquired internally during execution through sensor-reading operations and subsequently passed to the inference pipeline within $\f$.

\subsubsection*{\textbf{G3 (Low Overhead)}}
To accommodate \dev's limited resources, \acro must incur minimal overhead on \dev. 
We consider three types of overhead:

    \textbf{G3-1 (TCB):} The increase in Secure World TCB size must be small and independent of \f, allowing the same TCB to support arbitrary implementations of input retrieval and inference.

    \textbf{G3-2 (Memory):} \acro should minimize memory usage by avoiding duplication of code or model data across Secure and Normal Worlds whenever possible.

    \textbf{G3-3 (Latency):} \acro must minimize runtime overhead during inference.

\section{Design Overview}\label{sec:overview}

\subsection{Challenges}\label{subsec:challenges}

Simultaneously achieving \textbf{G1} to \textbf{G3} is challenging. 
To illustrate this, we focus on a subset of these: \textbf{G1-1} and \textbf{G3}. 
Even under this simplification, we show that commonly used designs in the literature -- \textbf{D1}, \textbf{D2}, and \textbf{D3} -- fail to satisfy both of them. 
Since \acro also targets more goals, it faces even more challenges.


{\textbf{D1: Full Normal World Execution.}}
We start with a strawman approach that executes \f entirely in \emph{Normal World}. 
While this approach naturally satisfies \textbf{G3} and is easy to deploy, it provides no security guarantees: \advMod can trivially violate \textbf{G1-1}, e.g., by extracting $\m$ from \dev's flash/RAM while \advInf can steal the attestation key in Normal World, resulting in a successful forgery of $\infToken$.

{\textbf{D2: Fully Secure World Execution.}}
At the other end of the spectrum, \f can be placed and executed entirely in the Secure World. 
To protect $\m$ at rest, \modOwner can provision it in encrypted form (during \textbf{I1: Provisioning}), which is then decrypted and loaded into secure RAM for execution by \f post-boot.
This satisfies \textbf{G1-1}, not \textbf{G3}. 
In particular, \textbf{G3-1 (TCB)} is violated, as the Secure World must include \f, which is application-specific (e.g., dependent on input sensors and inference implementations). 
Additionally, \textbf{G3-2 (Memory)} is strained, since \dev must store both an encrypted version of $\m$ in flash and a plaintext version in secure RAM. 
Finally, this design limits the use of emerging ML hardware accelerators (e.g., microNPUs~\cite{millar2025benchmarking}, Edge TPUs~\cite{yazdanbakhsh2021evaluation}), which are typically exposed to and managed by the Normal World~\cite{tfm_corstone320}, thereby hindering potential performance gains.

{\textbf{D3: Split Inference Across Worlds.}}
A middle-ground partitions $\m$ into private ($\mPriv$) and public components ($\mPub$). 
Here, $\mPriv$ is provisioned in encrypted form and executed in Secure World, similar to \textbf{D2}; $\mPub$ is managed and executed in the Normal World, as done in \textbf{D1}. 
This design is commonly used in prior TEE-based work~\cite{li2025teeslice,mo2020darknetz,11130497} to achieve \textbf{G1-1} on potentially untrusted devices.
Since $\mPriv$ is typically much smaller than $\mPub$, this design reduces secure memory requirements compared to \textbf{D2}, partially addressing \textbf{G3-2 (Memory)}. 
However, it does not resolve \textbf{G3-1 (TCB)}, as Secure World must still host \f to ensure correct usage of $\mPriv$. 
Moreover, \textbf{G3-3 (Latency)} can be negatively impacted due to frequent world switches between Secure and Normal Worlds, with overhead depending on the partitioning strategy, e.g., approaches like TEESlice~\cite{li2025teeslice}, which distribute private components across multiple layers, can incur substantial switching costs. 
Finally, this design still limits the use of emerging low-cost hardware accelerators for executing $\mPriv$.

\begin{table}[!htp]
\centering
\scriptsize
\caption{Memory access control of TrustZone-M domains with the introduction of \iobox; C and D indicate code and data, respectively, while PrivD and PubD refer to private and public data. RO indicates read-only access.}
\label{tab:access-control}
\begin{tabular}{l|cc|ccc|cc}
\toprule
 & \multicolumn{2}{c|}{NW} & \multicolumn{3}{c|}{\iobox} & \multicolumn{2}{c}{SW} \\
 & C & D & C & PrivD & PubD & C & D \\
\midrule
Normal World (NW) 
& \checkmark & \checkmark & RO & $\times$ & RO & $\times$ & $\times$ \\

\iobox 
& \checkmark & \checkmark & \checkmark & \checkmark & \checkmark & $\times$ & $\times$ \\

Secure World (SW)
& \checkmark & \checkmark & \checkmark & \checkmark & \checkmark & \checkmark & \checkmark \\
\bottomrule
\end{tabular}
\end{table}

\subsection{High-Level Intuition}

From \textbf{D1--D3}, we observe that \textbf{G3} is achieved only by \textbf{D1}.
This suggests that extending \textbf{D1} with stronger security mechanisms could enable us to achieve both \textbf{G1-1} and \textbf{G3}.

Guided by this insight, we propose a new abstraction, \iobox, which executes \f in Normal World, inheriting \textbf{D1}'s benefits for achieving \textbf{G3}. 
To enforce \textbf{G1-1}, we introduce a small, \f-agnostic component in the Secure World that elevates \iobox privilege during its execution, placing its privilege between Normal and Secure Worlds. 

With this new privilege, Normal World cannot access or tamper with \iobox's code and data, ensuring \textbf{G1-1}. 
At the same time, \iobox is prevented from interfering with Secure World. 
As a result, this design enables third-party developers to deploy isolated code without modifying highly trusted Secure World components.
%
Overall, \iobox establishes a new, third, runtime environment on TrustZone-M where 
the privilege relationships among the three environments are summarized in Table~\ref{tab:access-control}.

\textbf{Composition with Cryptographic Authorization and Proofs of Execution.}
To achieve complete protection, we compose \iobox to support \textbf{G1-2} and \textbf{G2} without sacrificing \textbf{G3}. Specifically, \iobox is extended with two features:
(1) an authorization mechanism that allows a trusted party to control who can invoke \iobox and how often, thereby enforcing \textbf{G1-2} (assuming $F$ correctly performs input retrieval and inference over $\m$), and
(2) a proof-of-execution mechanism that enables Secure World to produce verifiable evidence that isolated code within \iobox executed correctly and that a given output corresponds to that execution.

\section{\iobox Isolation Abstraction: A Third Runtime Environment on TrustZone-M}\label{sec:iobox}

Here, we describe \iobox, starting with its high-level lifecycle and then detailing the Secure-World APIs used to orchestrate its lifecycle state.

\subsection{Lifecycle of \iobox}

\begin{figure*}[t]
    \centering
    \begin{minipage}{0.35\textwidth}
        \centering
        \vspace{3em}
        \includegraphics[width=\linewidth]{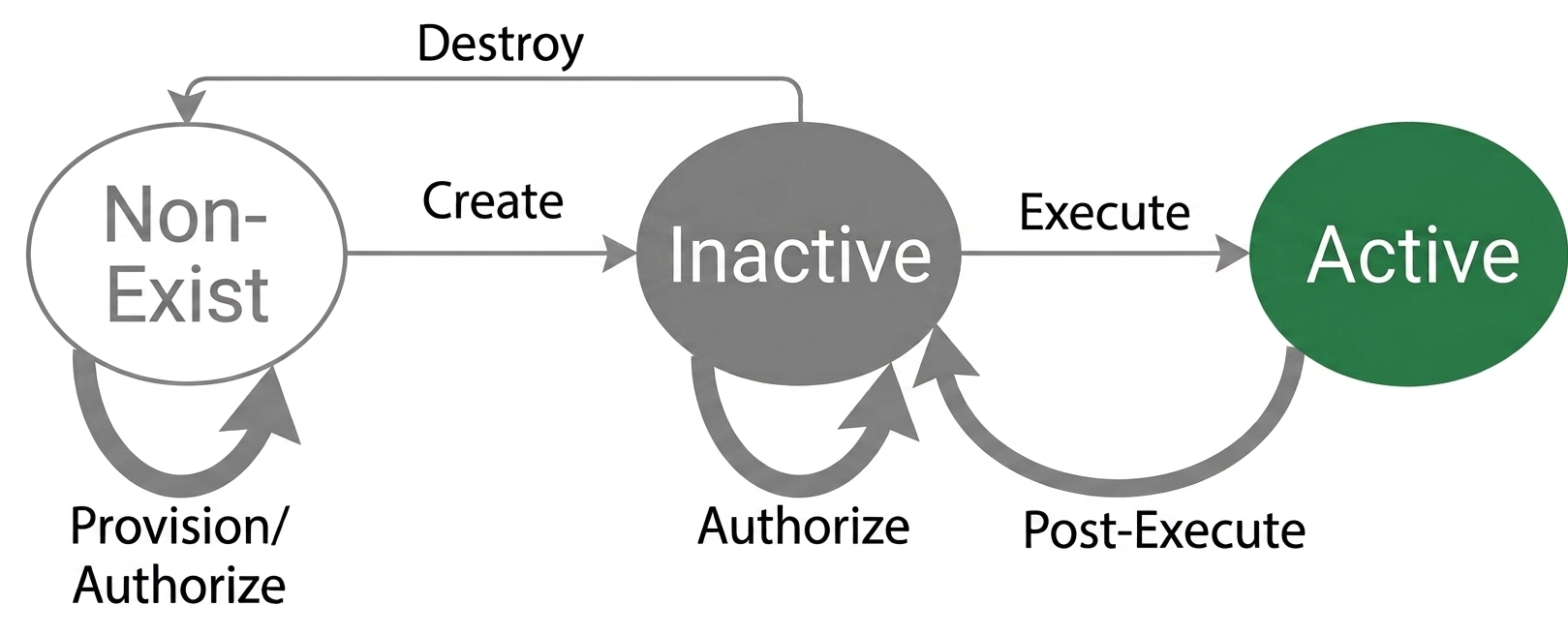}
        \subcaption{\iobox Lifecycle.}
        \label{fig:lifecycle}
    \end{minipage}
    \hfill
    \begin{minipage}{0.6\textwidth}
        \centering
        \includegraphics[width=\linewidth]{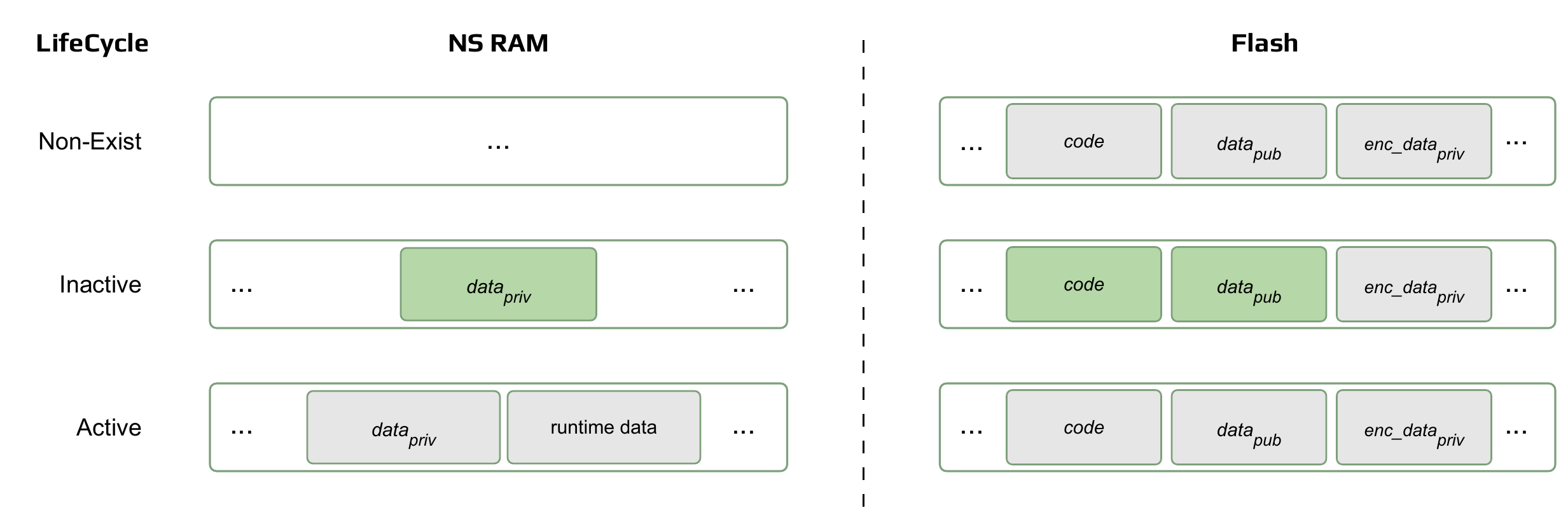}
        \subcaption{Gray denotes Non-Secure state; Green denotes Secure.}
        \label{fig:iobox-states}
    \end{minipage}
    \caption{Overview of \iobox lifecycle (a) and its corresponding TrustZone-M memory state (b).}
    \label{fig:iobox-overview}
\end{figure*}

The lifecycle of \iobox (Figure~\ref{fig:lifecycle}) consists of three states: \texttt{Non-Exist}, \texttt{Inactive}, and \texttt{Active}, with transitions controlled by four Secure World APIs.

On a new \dev, \iobox is in the \texttt{Non-Exist} state. 
During provisioning, \texttt{Provision} API is called to install the necessary \iobox metadata (e.g., code/data hash and keys), while \texttt{Authorize} configures access control policies such as permitted users and usage limits. 
These steps prepare \iobox for deployment but do not instantiate it, so it remains in the \texttt{Non-Exist} state.

To create an executable instance, the Normal World invokes \texttt{Create}. 
This validates \iobox's code and data with the stored metadata, initializes and assigns it the access privileges defined in Table~\ref{tab:access-control}, and transitions it to the \texttt{Inactive} state, where it is ready to run.
Invoking \texttt{Execute} transitions \iobox to the \texttt{Active} state, during which it executes $\f$. 
Upon completion, it returns to the \texttt{Inactive} state. 
This enables repeated executions by alternating between \texttt{Inactive} and \texttt{Active} via \texttt{Execute}.
While in the \texttt{Inactive} state, \texttt{Authorize} can be invoked to update \iobox authorization policy without requiring re-creation.

When the \iobox instance is no longer needed, \texttt{Destroy} can be invoked to erase its sensitive state and release its memory back to the Normal World, returning to the \texttt{Non-Exist} state.
Next, we describe how the aforementioned Secure World APIs are designed/implemented using TrustZone-M.


\begin{algorithm}[htp]
\scriptsize
\caption{\iobox Secure World APIs}
\label{alg:secure-apis}
\DontPrintSemicolon
\SetKwProg{Fn}{API}{}{}
\SetKwFunction{Verify}{Verify}
\SetKwFunction{Sign}{Sign}
\SetKwFunction{AuthDec}{AuthDec}
\SetKwFunction{Hash}{Hash}

\textbf{Context:} $CTX: H_{s\_id} \rightarrow ( k_{dec}, pk_{o}, pk_{u}, data\_id, entry, usage, limit, state)$

\vspace{2mm}

\Fn{\texttt{Provision}($k_{dec}, pk_{o}, pk_{u}, s\_id, limit$)}{
    $H_{s\_id} \leftarrow \Hash(s\_id)$\;
    parse $(data\_id, entry)$ from $s\_id$\;
    $CTX[H_{s\_id}] \leftarrow (k_{dec}, pk_{o}, pk_{u}, data\_id, entry, 0, limit, \textsf{Non-Exist})$
}

\vspace{2mm}


\vspace{2mm}

\Fn{\texttt{Authorize}($pk_{u}, limit, H_{s\_id}, T_{o}$)}{
    \If{$H_{s\_id} \notin CTX$}{abort\;}
    retrieve $pk_{o}$ from $CTX[H_{s\_id}]$\;
    \If{$limit \le CTX[H_{s\_id}].limit$ \textbf{or} \textbf{not} \Verify($pk_{o}$, $T_{o}$, $pk_{u} \| limit \| H_{s\_id}$)}{abort\;}
    $CTX[H_{s\_id}].\{pk_{u},limit\} \leftarrow \{pk_{u},limit\}$
}

\vspace{2mm}

\Fn{\texttt{Create}($s\_id$)}{
    $H_{s\_id} \leftarrow \Hash(s\_id)$\;
    \If{$H_{s\_id} \notin CTX$}{\Return failure\;}

    parse $(\f, data_{pub}, enc\_data_{priv})$ from $s\_id$\;

    mark $(\f, data_{pub}, data_{priv})$ as Secure\;

    $data_{priv} \leftarrow \mathsf{Dec}(CTX[H_{s\_id}].k_{dec}, enc\_data_{priv})$\;

    $CTX[H_{s\_id}].state \leftarrow \textsf{Inactive}$\; 
}

\vspace{2mm}

\Fn{\texttt{Execute}($u, In, H_{s\_id}, proof, T_{u}$)}{
    \If{$H_{s\_id} \notin CTX$}{abort\;}

    $(pk_o, pk_{u}, data\_id, entry, usage, limit, state) \leftarrow CTX[H_{s\_id}]$\;

    \If{$state \neq \textsf{Inactive}$ \textbf{or} $usage \ge limit$ \textbf{or} $u \leq usage$}{abort\;}
    \If{\textbf{not} \Verify($pk_{u}$, $T_{u}$, $u \parallel In \parallel H_{s\_id} \parallel proof$)}{abort\;}

    disable interrupts\;

    $CTX[H_{s\_id}].state \leftarrow \textsf{Active}$\;

    allocate \iobox stack\;
    mark $(\f, data_{pub}, data_{priv})$ as Non-secure\;

    $Out_\f \leftarrow$ execute $entry(In)$\;

    erase stack\;
    mark $(\f, data_{pub}, data_{priv})$ as Secure\;

    $CTX[H_{s\_id}].\{usage,state\} \leftarrow \{u,\textsf{Inactive}\}$\; 

    enable interrupts\;

    \If{$proof$}{
        $\infToken \leftarrow \Sign(sk_{\dev}, \f \parallel u \parallel data\_id \parallel \fIn \parallel \fOut \parallel pk_o)$\;
    }

    \Return $(\fOut, \infToken)$\;
}

\vspace{2mm}

\Fn{\texttt{Destroy}($H_{s\_id}$)}{
    \If{$H_{s\_id} \notin CTX$}{abort\;}

    erase $data_{priv}$\;
    mark $(\f, data_{pub}, data_{priv})$ as Non-secure\;

    $CTX[H_{s\_id}].state \leftarrow \textsf{Non-Exist}$\;

}

\end{algorithm}

\subsection{Secure-World APIs}

To orchestrate the \iobox lifecycle, we maintain its context in secure persistent storage and expose a set of Secure-World APIs that allow the Normal World to trigger controlled state transitions without directly accessing or manipulating \iobox code and data.

As a basic form, we provide \texttt{Provision}, \texttt{Create}, \texttt{Execute}, and \texttt{Destroy}, which together ensure that \iobox adheres to the privilege model in Table~\ref{tab:access-control}.
We further introduce two optional extensions:
(1) an \texttt{Authorize} API that allows \dev owner to specify which users may access \iobox and how many times it can be invoked, and
(2) an input flag to \texttt{Execute} that enables generation of a proof of execution, allowing an external verifier to validate the authenticity of the output.

All APIs are summarized in Algorithm~\ref{alg:secure-apis}, while the corresponding TrustZone-M security states of \iobox memory across its lifecycle are illustrated in Figure~\ref{fig:iobox-states}.

\subsubsection{\texttt{Provision} (Line 2-5)}

Unlike the other APIs, \texttt{Provision} is only accessible to a trusted entity, e.g., during device manufacturing or via a Secure-World update mechanism that verifies the integrity of the \iobox configuration prior to its installation.

It takes as input the \iobox owner's public key $pk_o$ and a public identifier $s\_id$ describing the \iobox instance, containing its code \f, entry point $entry$, and a data identifier $d\_id$ that captures both public data $data_{pub}$ and encrypted private data $enc\_data_{priv}$ (if any).
If $enc\_data_{priv}$ is present, a decryption key $k_{dec}$ is also provided.
Optionally, the API accepts a user public key $pk_u$ and an access limit $limit$, specifying which user is authorized to invoke \f (which executes within \iobox) and how many times it may be executed.

Upon invocation, it computes a hash of $s\_id$ to serve as a unique identifier for the \iobox instance, initializes its Secure-World context ($CTX$) with the provided input, sets the instance's usage counter to $0$, and places the instance in the \texttt{Non-Exist} state.

\subsubsection{\texttt{Authorize} (Line 6-12)} This API updates the access policy of a provisioned \iobox.
It takes as input a user public key $pk_u$, a new limit $limit$, a unique \iobox identifier $H_{s\_id}$, and an authorization token $T_{o}$ over $(pk_u, limit, H_{s\_id})$ produced by the \iobox owner.
Upon invocation, it verifies that a \iobox instance identified by $H_{s\_id}$ has been provisioned and validates $T_{o}$ using $pk_o$.
If both checks succeed, it updates the instance's context with $pk_u$ and $limit$ accordingly.
This API enables the \iobox owner to dynamically change the authorized user and adjust the number of allowed invocations for that user.

\subsubsection{\texttt{Create} (Line 13-20)}
This API initializes a \iobox instance from its provisioned configuration, taking $s\_id$ as input.
Then, it retrieves the memory regions corresponding to the code \f, public data $data_{pub}$, and encrypted private data $enc\_data_{priv}$ from the instance's context.
It then uses the TrustZone-M's SAU and security DMA controller (recall Section~\ref{sec:bg} and~\ref{subsec:system-threat-model}) to ephemerally mark the code and $data_{pub}$ regions as Secure, protecting them from both Normal World access and Non-Secure DMA.
If encrypted private data exists, the API allocates a region $data_{priv}$, ephemerally marks it as Secure, and decrypts $enc\_data_{priv}$ using $k_{dec}$ into this region.
To prevent unauthorized reads by external peripherals, the API places $data_{priv}$ in a Non-Secure RAM region that does not overlap with any memory-mapped peripheral regions.
Finally, it sets the \iobox lifecycle state to \texttt{Inactive}, indicating that the instance has been populated with code and data and is ready for execution.
As illustrated in Figure~\ref{fig:iobox-states}, after this call, all \iobox code and data are in the Secure state, preventing access or modification by the Normal World.

\noindent \textbf{Remark.} Although \f and its data are temporarily marked as Secure, \f is never executed in this state. Instead, it is atomically restored to Non-Secure immediately before execution (see \texttt{Execute} below), ensuring that the \iobox state cannot be altered between its definition and execution and at the same time maintaining isolation between \f and the Secure World TCB.

\subsubsection{\texttt{Execute} (Line 21-40)} This API performs a single \f invocation in \iobox.
It takes as input a user message containing the post-execution usage counter $u$, \f input $In_\f$, $H(s\_id)$, and a signed token $T_{u}$ over this message produced by the user.

Upon invocation, it looks up a \iobox instance from $H_{s\_id}$ and retrieves its context.
It further checks that: (1) the instance is in the \texttt{Inactive} state, (2) the provided counter $u$ is consistent with the current usage (i.e., $u \geq usage + 1$), (3) the invocation does not exceed $limit$, and (4) authenticity of $T_{u}$ using $pk_u$.

If all checks succeed, Secure World prepares execution of \f in \iobox as follows:
it disables interrupts, sets the \iobox state to \texttt{Active}, and allocates a dedicated stack at a fixed location (that does not overlap with memory-mapped peripheral regions) in Normal World. 
Next, Secure World marks \f, $data_{pub}$ and $data_{priv}$ as Non-Secure (see Figure~\ref{fig:iobox-states}) and calls \f entry point ($entry$) to execute \f with input $In_\f$ in the Normal World.
Once \f execution completes, control returns to Secure World (this is achieved by requiring an NSC at the end of \f).
The Secure World then erases the \iobox stack, restores \iobox code and data to Secure state, increments the usage counter, and sets the lifecycle state back to \texttt{Inactive} before re-enabling interrupts.

If a proof of execution is required, Secure World generates \infToken by signing $(\f, u, data\_id, \fIn, \fOut)$ using the \dev-bound attestation key $sk_{\dev}$.
Finally, it returns $(\infToken, \fOut)$ to the caller.

\subsubsection{\texttt{Destroy} (Line 41-46)} This API tears down a \iobox instance $H_{s\_id}$.
Upon invocation, Secure World erases all sensitive data in $data_{priv}$.
It then marks \f, $data_{pub}$ and $data_{priv}$ regions as Non-Secure, releasing the memory back to the Normal World.
Finally, it sets the \iobox lifecycle state to \texttt{Non-Exist}.

\subsection{\iobox Security Analysis}\label{subsec:prop}

We argue that \iobox isolation abstraction achieves the following security properties:

\subsubsection*{\textbf{P1: Privilege Separation}}
We show that the four core APIs -- \texttt{Provision}, \texttt{Create}, \texttt{Execute} (without proof-of-execution), and \texttt{Destroy} -- are sufficient to enforce the privilege model in Table~\ref{tab:access-control}.

During \texttt{Execute}, \f runs atomically with interrupts disabled, and control is not returned to Normal World until both \f and \texttt{Execute} complete.
This prevents Normal World from interrupting execution to access $data_{pub}$ or $data_{priv}$ (which are in Non-Secure state) while \iobox is in the \texttt{Active} state.

A compromised Normal World may attempt to access \iobox code and data outside the \texttt{Active} state.
When \iobox is in the \texttt{Inactive} state, such access is prevented by the SAU, as all \iobox memory regions are marked Secure (Figure~\ref{fig:iobox-states}).
DMA-based attacks are mitigated in the same way, since the Secure/Non-Secure memory attributes also apply to DMA accesses.
The remaining opportunity is when \iobox is in the \texttt{Non-Exist} state.
In this state, \f and $data_{pub}$ reside in Non-Secure flash and may be read or modified.
However, any tampering is detected during \texttt{Create}, as the recomputed $H_{s\_id}$ (over \f and $data_{pub}$) will not match the provisioned value, preventing \iobox from being created.
For $data_{priv}$, confidentiality is preserved since it exists only in encrypted form in the \texttt{Non-Exist} state.

A subtle case arises if the device resets while \iobox is in the \texttt{Active} or \texttt{Inactive} state, where $data_{priv}$ or its derived runtime state may persist in Non-Secure RAM.
To address this, we propose either: (i) extending Secure-World initialization to clear all Non-Secure RAM at boot, or (ii) having \texttt{Execute} record the stack region in its context and extending Secure-World initialization to detect incomplete executions and securely erase the corresponding memory.

We note that \iobox always lives/executes in the Non-Secure state and therefore cannot access or interfere with higher-assurance Secure-World components. \f code and data are only marked as Secure in between \texttt{Create} and \texttt{Execute}, but never executed in this state, being brought back to Non-Secure upon \texttt{Execute} invocation. This dynamic remarking indeed is what enables \iobox existence and an intermediate security state between Normal and Secure Worlds.
At the same time, executing in the Non-Secure state allows \iobox to access and reuse existing Non-Secure code when needed.
As a result, these satisfy the privilege model in Table~\ref{tab:access-control}.

\subsubsection*{\textbf{P2: Authorization of \iobox Invocation}}
We show that, with the addition of the \texttt{Authorize} API, only authorized users can invoke \f in \iobox, and they cannot exceed the usage limit specified by the owner.

An unauthorized user cannot execute \f because their public key is not registered in the \iobox (Secure World-protected) context.
The \texttt{Execute} API requires a user message signed by the authorized user and verifies it against the authorized user's public key $pk_u$.
Without access to the corresponding private key, an adversary cannot produce a valid signature and thus cannot run a \iobox instance.

Furthermore, \texttt{Authorize} updates the authorized user and usage limit only upon receiving a valid owner-signed message (verified using $pk_o$).
The usage counter is restricted to Secure World and checked during \texttt{Execute} to ensure that each invocation is fresh and does not exceed the specified limit.
This prevents both overuse and replay attacks.

\subsubsection*{\textbf{P3: PoX Unforgeability}}
Finally, when the proof-of-execution feature is enabled ($proof = 1$ in \texttt{Execute}), we show that a valid \infToken can only be produced if \f is executed correctly within \iobox.
We adopt the notion of correct execution from prior work~\cite{nunes2020apex}, where \f must execute completely from its entry point to its exit point without interruption and the value of $\fOut$ must be determined by this execution.

\infToken is generated inside Secure World via the \texttt{Execute} API using $sk_{\dev}$ and covers $(\f, u, data\_id, \fIn, \fOut)$.
Since $sk_{\dev}$ is not accessible to the Normal World, an adversary cannot forge a valid \infToken.
Consequently, a valid \infToken can only be obtained by successfully invoking \texttt{Execute}, which runs \f with input $In_\f$ inside \iobox to produce the corresponding output $Out_\f$.
Any attempt to modify $In_\f$ or $Out_\f$ results in an invalid proof, as both are cryptographically bound in \infToken.
Also, interrupting \f execution to induce an incomplete execution is not possible due to interrupts being disabled during \f execution.
A reset during \texttt{Execute} may cause \f to terminate prematurely; however, in this case, no \infToken is produced, since it is generated only upon successful completion of \texttt{Execute}.
In all cases, an adversary cannot forge \infToken without having to execute \f inside \iobox with the correct $In_\f$ and produced $\fOut$.


\section{\acro: Edge Inference Protocol}
\label{sec:savid-interactions}

Building on the \iobox abstraction and its Secure-World APIs (Section~\ref{sec:iobox}), we present \acro protocol realizing interactions \textbf{I1--I4} defined in Section~\ref{sec:prelim}.

\subsection{I1--I4 in \acro}

\textbf{I1: Provisioning}
\modOwner prepares the software \f with entry point $entry$. 
\f is a self-contained binary that bundles the input-retrieval code (e.g., a sensor driver), the inference library, and the glue code connecting them, executing both back-to-back, i.e., it first retrieves the inference input directly from \dev's hardware peripherals, then immediately runs DNN inference on it. 
Given a DNN model $\m$, \modOwner assigns its ID as $\m_{ID}$ and then applies a TSDP partitioning strategy to split it into $\mPriv$ and $\mPub$.
It then generates $k_{dec}$ and uses it to encrypt $\mPriv$.

Using $s\_id$, \modOwner installs $\f$, $\mPub$, and encrypted $\mPriv$ in the Normal World of \dev, 
while provisioning the corresponding \iobox metadata to the Secure World via \texttt{Provision} API.
Besides $s\_id$ and $k_{dec}$, this metadata also includes \modOwner's public key $pk_{\modOwner}$, 
and optionally the authorized user's public key $pk_{\modUser}$ and the usage $limit$.

Note that \acro is generic and supports both using and not using TSDP. In the latter case $\mPriv$ is the entire model and $\mPub$ is empty.

\textbf{I2: Authorization}
\modUser requests permission from \modOwner to invoke inference on \dev.
If approved, \modOwner issues a signed authorization token $\authToken$ that grants the user with public key $pk_{\modUser}$ access to the \iobox instance identified by $H_{s\_id}$ with a usage limit $limit$.

\modUser forwards $\authToken$ to \dev, which invokes the \texttt{Authorize} API to update the access policy of the corresponding \iobox instance.
Upon successful verification of $\authToken$, the Secure World registers $pk_{\modUser}$ and $limit$ for $H_{s\_id}$.

\textbf{I3: Inference}
To perform inference, \modUser first invokes the \texttt{Create} API to instantiate the \iobox instance identified by $H_{s\_id}$.
It then constructs a request containing the next usage counter $u$, the input $In_\f$, the identifier $H_{s\_id}$, and a flag $proof = 1$ indicating that a proof of execution is required.
This request is signed using $sk_{\modUser}$ to produce an inference token $T_{u}$.

\modUser submits the request and $T_{u}$ to the \texttt{Execute} API.
Upon successful verification, \texttt{Execute} runs the corresponding function $\f$ within \iobox, which retrieves input data (e.g., from a sensor) and performs inference using $\m$, producing an output $Out_\f$.
Since $proof = 1$, this invocation returns a proof of execution $\infToken$.

\textbf{I4: Proof Verification}
\modUser forwards $(\fOut, \infToken)$ to \vrf.
\vrf checks that $\infToken$ is a valid token under $pk_{\dev}$ over the expected $pk_{\modOwner}$, $u$, $\f$, $\m_{ID}$, $In_\f$, and $\fOut$.
%

\subsection{\acro Security Analysis}\label{subsec:acro}

We show that \acro satisfies \textbf{G1} and \textbf{G2}. 
Our arguments build on the security properties of \iobox established in Section~\ref{subsec:prop}.

\subsubsection*{\textbf{G1-1 (Direct Leakage Protection)}}
The DNN model $\m$ is partitioned into $\mPub$ and $\mPriv$, where $\mPub$ is deliberately set as public by \modOwner.
Thus, \textbf{G1-1} concerns only the protection of $\mPriv$.

In \acro, $\mPriv$ is treated as private data within \iobox and therefore inherits \textbf{P1: Privilege Separation} that guarantees the access control guarantees in Table~\ref{tab:access-control}.
In particular, untrusted Normal World on \dev cannot access $\mPriv$ in plaintext.
Outside \dev, the only point at which \mPriv is handled in plaintext is during \textbf{I1: Provisioning} by \modOwner.
As \advMod excludes \modOwner, \advMod cannot directly observe $\mPriv$, satisfying \textbf{G1-1}.

\subsubsection*{\textbf{G1-2 (Indirect Leakage Protection)}}
Indirect leakage through overused or unauthorized inference queries is mitigated through \textbf{I2: Authorization} in \acro.
Therein, \modUser must obtain a valid $\authToken$ from \modOwner, which will be enforced on \dev by the \texttt{Authorize} API before any inference can be executed through \texttt{Execute}.
As a result, \acro inherits \textbf{P2: Authorization of \iobox Invocation}, preventing unauthorized \advMod from performing inference on \m more than \modOwner's specified usage limit.

\subsubsection*{\textbf{G2-1 (Authenticity)}}

In \acro, \textbf{I3: Inference} requires invoking the \texttt{Execute} API, thereby inheriting \textbf{P3: PoX Unforgeability} of \iobox.
In particular, any attempt by $\advInf$ to forge an inference result $Out_\f$ or its corresponding proof $\infToken$ will be detected by \vrf.
Freshness is ensured through the inclusion of $u$ in \infToken, where $u$ serves as a nonce to prevent replay attacks.
\vrf can obtain an up-to-date value of $u$ from a trusted source, such as \modOwner.

Moreover, since $\f$ encapsulates both input retrieval and inference, a valid $\infToken$ guarantees that the inference input is obtained without being tampered with and that $\f$ executes completely without interruption. The acquisition of local \dev sensor inputs is implemented in \f code and thus implicitly checked by checking \f proper execution.
As $\infToken$ also covers authentic $\m_{ID}$ and $pk_\modOwner$ (protected on \dev through Secure World's secure storage), \vrf can determine whether the inference is performed using the right DNN model $\m$.
In sum, \acro assures that $Out_\f$ is the authentic result of executing $\f$ on input $In_\f$ using the model identified by $\m_{ID}$, as provisioned by \modOwner. Thus, \textbf{G2-1} is fulfilled.

\subsubsection*{\textbf{G2-2 (Privacy)}}
In \textbf{I4: Proof Verification}, verification relies on public information, including $\f$, $\fIn$, $\m_{ID}$, $pk_{\modOwner}$, $\fOut$ and $\infToken$.
Importantly, the raw inference input is neither revealed nor required during verification, as it is acquired and processed internally within \iobox (also not exposed to \dev's Normal World).
Hence, \acro preserves \dev raw inputs privacy, satisfying \textbf{G2-2}.

\section{Evaluation}\label{sec:eval}

We implement \acro on a NUCLEO-L552ZE-Q development board~\cite{st_nucleo_l552ze_q} with an ARM Cortex-M33 MCU, 192\,KB of Normal-World SRAM (NS RAM), and 64\,KB of Secure-World SRAM (S RAM).
For DNN inference, we use ResNet~\cite{targ2016resnet}, which is commonly used in low-end edge inference studies~\cite{11130497,ghosh2023energy,liu2025ed}. Following~\cite{11130497}, we partition the model into \mPub and \mPriv by placing the last three layers in \mPriv. This yields $|\m|=76.375$\,KB, with $|\mPriv|=38.625$\,KB and $|\mPub|=37.75$\,KB.
We perform \textbf{I1: Provisioning} offline by manually installing \f, emulating secure factory provisioning; hence, provisioning time is excluded from our measurements. Additional implementation details are provided in Appendix~\ref{apdx:impl-details}, and limitations/extensions are discussed in Appendix~\ref{apdx:limit}.

\textbf{Baseline.}
We compare \acro against a baseline where \f and inference-result generation execute entirely in the Secure World. To provide comparable security guarantees, the baseline authenticates inference requests and signs inference outputs using a \dev-bound key. This baseline corresponds to \textbf{D2} in Section~\ref{subsec:challenges}.

\begin{table}[t]
\scriptsize
\centering
\caption{TCB comparison: \acro vs the baseline.}
\label{tab:tcb_comparison}
\begin{tabular}{lcc}
\toprule
\textbf{\acro API} & \textbf{LOC} & \textbf{\% of Baseline (34,865 LOC)} \\
\midrule
\texttt{Authorize} & 264 & 0.76\% \\
\texttt{Create}    & 576 & 1.65\% \\
\texttt{Execute}   & 451  & 1.29\% \\
\texttt{Destroy}   & 232 & 0.66\% \\
\midrule
\textbf{Total TCB} & \textbf{1523} & \textbf{4.36\%} \\
\bottomrule
\end{tabular}
\end{table}

\subsection{Results for G3-1 (TCB)}
The baseline runs \f in the Secure World, so the TCB includes \f, which is dominated by the CMSIS-NN library. The remaining code handles request authentication and output signing, which are negligible compared to the library.

Table~\ref{tab:tcb_comparison} reports the Secure-World TCB of \dev, broken down by component and compared to the baseline. As expected, \texttt{Create} and \texttt{Execute} have larger TCB contribution due to validation checks, memory management, and cryptographic operations. In contrast, \texttt{Destroy} has the smallest TCB, as it performs a simple operation of memory erasure.

Compared to the baseline, \acro reduces the TCB from 34,865 LOC to 1523 LOC (4.36\% of the baseline).
More importantly, the TCB on \dev is agnostic to \f, and hence to sensor reading driver and inference library, making it unchanged across different combinations of sensor-input reading implementation and model architectures/DNN library (e.g., TensorFlow-Lite~\cite{david2021tensorflow}). 
This contrasts with the baseline approach, where the TCB grows with a new \f function, making it hard to realize for multiple \f variants. This satisfies \textbf{G3-1 (TCB)}.

\subsection{Results for G3-2 (Memory)}

In the baseline, the full model $\m$ executes in the Secure World and is treated as confidential. To prevent extraction from Flash (Flash dump), $\m$ must be stored encrypted and decrypted into S RAM at runtime, creating two copies: ciphertext in Flash and plaintext in S RAM. In the test case, this requires 76.375~KB of S RAM (does not actually fit into the NUCLEO-L552ZE-Q available S RAM).

In contrast, \acro partitions $\m$ into $\mPub$ and $\mPriv$.
Only $\mPriv$ is stored encrypted in Flash, while $\mPub$ remains in plaintext and can be accessed directly from Flash at runtime.
During execution, \acro instantiates \iobox that decrypts $\mPriv$ into the protected NS RAM.
Thus, only $\mPriv$ is duplicated, whereas $\mPub$ exists as a single Flash-resident copy.

As a result, \acro substantially reduces runtime memory duplication and moves the duplicated memory footprint away from S RAM.
Unlike the baseline, the duplicated model portion $\mPriv$ resides in NS RAM (securely protected by \iobox) rather than in S RAM, making more Secure-World memory available to other security-critical services.
Also, since \acro is agnostic to the partitioning strategy, deployments can further reduce memory overhead by selecting strategies that minimize the size of $\mPriv$. This satisfies \textbf{G3-2 (Memory)}.

\subsection{Results for G3-3 (Latency)}
\label{subsec:Latency}
\begin{table}[t]
\scriptsize
\centering
\caption{Inference latency of \acro vs. baseline.} 
\label{tab:benchmark}
\begin{tabular}{lccc}
\toprule
NS baseline & \acro & Overhead \\
\midrule
 1160.93 ms & 1161.76 ms & 0.83 ms (0.07\%)\\
\bottomrule
\end{tabular}
\end{table}

For \textbf{G3-3 (Latency)}, we measure the end-to-end inference latency of \texttt{Execute}. All results reported are obtained as the average over 100 executions.
For the baseline, the full decrypted model $\m$ requires about 76.375~KB.  However, S RAM is limited to 64~KB, making the baseline infeasible for direct comparison.
Instead, we use an \emph{NS baseline}, which mimics the original baseline design but loaded in the Normal World, where a larger RAM size is available (192~KB).


Table~\ref{tab:benchmark} reports the inference latency of \acro (via the \texttt{Execute} API) and the NS baseline. Compared to the baseline, \acro adds only 0.83~ms of latency, for a total runtime of 1161.76~ms. 
The overhead includes memory region reconfiguration, \iobox stack setup and erasure, and state management. 
The PoX computation adds an additional 230.38~ms to generate the cryptographic signature/proof sent to \vrf. 
However, in the baseline, the inference output must also be signed by the Secure World (to at least authenticate \dev), resulting in the same cost also applying to the NS baseline.
The runtimes of the remaining operations are shown in Table~\ref{tab:api_latency}. 

As the runtime of \texttt{Create} and \texttt{Destroy} depends on the size of $\mPriv$, we further evaluate their runtime across varying $\mPriv$ sizes.
Figure~\ref{fig:create_latency} shows that \texttt{Create} latency scales linearly with $\mPriv$, increasing from 19.32~ms at 5~KB to 76.57~ms at 39.5~KB.
By comparison, \texttt{Destroy} remains substantially faster (Figure~\ref{fig:destroy_latency}), ranging from 0.448~ms to 2.639~ms across the same configurations.
In practical deployments, \iobox construction and destruction would likely occur once per session rather than once per inference, making their amortized overhead small.

\begin{figure}
\centering

\begin{subfigure}[t]{0.48\linewidth}
\centering
\begin{tikzpicture}
\begin{axis}[
    width=\linewidth,
    height=3.2cm,
    scaled x ticks=false,
    grid=major,
    grid style={dashed, gray!30},
    xlabel={$\mPriv$ size (KB)},
    ylabel={Latency (ms)},
    ymin=0,
    ymax=90,
    xtick={5120, 10240, 15360, 20480, 25600, 30720, 35840, 39552},
    xticklabels={5,10,15,20,25,30,35,39},
    tick label style={font=\scriptsize},
    label style={font=\scriptsize},
    axis lines=left,
    line width=0.8pt,
]

\addplot[
    thick,
    mark=*,
    mark size=1.5pt,
] coordinates {
    (5120, 19.326)
    (10240, 27.821)
    (15360, 36.341)
    (20480, 44.836)
    (25600, 53.357)
    (30720, 61.852)
    (35840, 70.371)
    (39552, 76.572)
};

\end{axis}
\end{tikzpicture}
\caption{\texttt{Create}}
\label{fig:create_latency}
\end{subfigure}
\hfill
\begin{subfigure}[t]{0.48\linewidth}
\centering
\begin{tikzpicture}
\begin{axis}[
    width=\linewidth,
    height=3.2cm,
    scaled x ticks=false,
    grid=major,
    grid style={dashed, gray!30},
    xlabel={$\mPriv$ size (KB)},
    ylabel={Latency (ms)},
    ymin=0,
    ymax=3.0,
    xtick={5120, 10240, 15360, 20480, 25600, 30720, 35840, 39552},
    xticklabels={5,10,15,20,25,30,35,39},
    tick label style={font=\scriptsize},
    label style={font=\scriptsize},
    axis lines=left,
    line width=0.8pt,
]

\addplot[
    thick,
    mark=square*,
    mark size=1.5pt,
] coordinates {
    (5120, 0.448)
    (10240, 0.774)
    (15360, 1.100)
    (20480, 1.427)
    (25600, 1.751)
    (30720, 2.079)
    (35840, 2.406)
    (39552, 2.639)
};

\end{axis}
\end{tikzpicture}
\caption{\texttt{Destroy}}
\label{fig:destroy_latency}
\end{subfigure}

\caption{Latency of \texttt{Create} and \texttt{Destroy}.}
\label{fig:lifecycle_latency}
\end{figure}

\begin{table}[t]
\centering
\scriptsize
\caption{\acro latency breakdown.}
\label{tab:api_latency}
\begin{tabular}{lrrrr}
\toprule
\textbf{Metric} & \texttt{Authorize} & \texttt{Create} & \texttt{Execute} & \texttt{Destroy} \\
\midrule
\textbf{Cycles} & 50,273,157 & 8,423,049 & 127,793,297 & 290,201   \\
\textbf{Latency (ms)} & 457.03 & 76.57 & 1161.76 & 2.64  \\
\bottomrule
\end{tabular}
\end{table}

\subsection{Case Study}\label{subsec:case-study}

To validate \acro in a realistic deployment setting, we implement a complete image
classification pipeline on the NUCLEO-L552ZE-Q board (including image acquisition via UART from a hardware camera interface).
\dev receives images over a UART-based camera stream, runs inference using the encrypted ResNet model, and returns classification results.
This pipeline is included in our open-source repository.

The case study follows \acro protocol in Section~\ref{subsec:acro}:
During \textbf{I1: Provisioning}, \modOwner partitions $\m$ and provisions both the model and $\f$ onto \dev in an offline phase.
Next, in \textbf{I2: Authorization}, \modUser obtains authorization from \modOwner and forwards the resulting $\authToken$ to \dev, which updates the access policy with a bounded inference budget.
During \textbf{I3: Inference}, \modUser creates the corresponding \iobox instance and issues inference requests within the authorized budget.
Once the budget is exhausted, the \iobox instance is destroyed.
We then verify inference outcomes in \textbf{I4: Proof Verification}.
Across 10 sequential inference executions, \acro incurs an average latency of 1428.42 ms per inference compared to 1427.57 ms for the NS baseline, indicating negligible runtime overhead.

\textbf{Security tests.}
We further validate \acro against attacker actions, consistent
with our threat model, by executing one directed negative test per attack type:

\textit{Memory Access Attacks.} We attempt to read the \iobox protected RAM region holding \mPriv from Non-Secure code, and similarly attempt to write protected Flash regions mapped to \iobox code.
All cases trigger hardware faults and are blocked by SAU,
confirming correct memory isolation.

\textit{Protocol Attacks.}
We test three protocol violations: (i) replaying a previously valid inference
request, (ii) issuing an inference request after budget exhaustion, and
(iii) forging the session-model binding $H_{s\_id}$.
All are rejected without modifying Secure state.

These tests confirm that \acro enforces authorization, freshness, and
model-identity binding per goals introduced in Section~\ref{sec:goals}.
Combined with the low latency overhead (Section~\ref{subsec:Latency}),
the reduced Secure memory footprint, and the
model-independent Secure TCB (Table~\ref{tab:tcb_comparison}), these results show that \acro is
a viable and secure approach for resource-constrained Cortex-M devices.

\begin{table}
\centering
\small
\caption{Comparison with prior TrustZone (TZ)-based work on edge inference (\pmark\ means partially-fulfilled).}
\label{tab:related}
\setlength{\tabcolsep}{5pt}
\scriptsize
\resizebox{\columnwidth}{!}{%
\begin{tabular}{lcccccccc}
\toprule
 & \textbf{TEE} & \textbf{G1-1} & \textbf{G1-2} & \textbf{G2-1} & \textbf{G2-2}
       & \textbf{G3-1} & \textbf{G3-2} & \textbf{G3-3} \\
\midrule
Darknetz~\cite{mo2020darknetz}      & TZ-A & $\checkmark$ & \pmark & $\times$ & $\times$ & $\times$ & $\checkmark$ & $\checkmark$ \\
ShadowNet~\cite{sun2023shadownet}   & TZ-A & $\checkmark$ & $\times$ & $\times$ & $\times$ & $\times$ & $\checkmark$ & $\times$ \\
ASGARD~\cite{moon2025asgard}        & TZ-A & $\checkmark$ & $\times$ & $\times$ & $\times$ & $\times$ & $\times$ & $\times$ \\
SecureQNN~\cite{11130497}     & TZ-M & $\checkmark$ & \pmark & $\times$ & $\times$ & $\times$ & $\checkmark$ & $\checkmark$ \\
\midrule
\textbf{\acro}               & TZ-M & $\checkmark$ & $\checkmark$ & $\checkmark$ & $\checkmark$ & $\checkmark$ & $\checkmark$ & $\checkmark$ \\
\bottomrule
\end{tabular}
}
\end{table}

\section{Related Work}


\textbf{Secure DNN Inference with TEEs.}
Recent work uses TEEs to secure DNN inference on edge devices. One approach executes the entire inference pipeline inside a TEE and uses attestation for verifiability~\cite{volos2018graviton,duddu2024laminator,chantasantitam2026pal,bayerl2020offline}. However, these systems rely on higher-end hardware, such as custom GPUs~\cite{volos2018graviton}, Intel SGX~\cite{duddu2024laminator}, or NVIDIA H100~\cite{chantasantitam2026pal}, making them unsuitable for the low-cost devices targeted in this work.

A lighter alternative is \emph{TEE-shielded DNN partitioning} (TSDP)~\cite{li2025teeslice}, which splits inference between protected layers inside the TEE and unprotected layers outside it. Prior work protects the exposed computation either by obfuscating layers/activations before offloading~\cite{hou2021model,tramer2018slalom,zhang2024verisplit}, or by selecting private layers to execute inside the TEE based on efficiency~\cite{li2025teeslice,moon2025asgard,10949698} or leakage risks~\cite{mo2020darknetz,sun2023shadownet,11130497}.

However, both full-TEE inference and TSDP-based designs require inference-related code to reside inside the trusted environment. On TrustZone-based systems, this enlarges the Secure-World TCB and increases the attack surface. In contrast, \acro executes inference code in \iobox, at the same privilege level as untrusted software but shielded from it, avoiding inference-code inclusion in the Secure World. Since \acro does not modify the model architecture or inference pipeline, it remains compatible with existing TSDP strategies. Table~\ref{tab:related} compares \acro with prior TrustZone-based systems.

\textbf{Verifiable Execution and Isolation on Edge Devices.}
Our \iobox abstraction resembles enclave-/realm-based execution on higher-end platforms such as Intel SGX and ARM CCA. However, TrustZone does not natively provide such an abstraction for executing security-sensitive code in the Normal World while isolating it from other Normal-World software; this is especially challenging on TrustZone-M, which lacks virtual memory and user-space isolation.

Prior work brings enclave-style execution to TrustZone-A~\cite{feng2021scalable,brasser2019sanctuary}, but these designs rely on hardware features unavailable on low-end TrustZone-M devices: Penglai~\cite{feng2021scalable} requires an MMU, while Sanctuary~\cite{brasser2019sanctuary} assumes multi-core platforms. Other low-end-device mechanisms provide specific services, such as software integrity verification~\cite{eldefrawy2012smart,nunes2019vrased,noorman2013sancus,nunes2024toward}, verifiable function execution~\cite{rattanavipanon2025slapp,nunes2020apex,neto2025pearts}, secure software updates~\cite{de2022casu,de2019pure}, confidential sensor input protection~\cite{nunes2022privacy,de2024sa4p}, or compartmentalization~\cite{ucca}. 
While these approaches provide important building blocks, none of them achieves the full set of goals targeted in this work by themselves.

\ignore{
\textbf{Verifiable Execution and Isolation on Edge Devices.}
Our \iobox abstraction resembles enclave/realm-based execution models found in higher-end platforms such as Intel SGX enclaves or ARM CCA Realms.
However, ARM TrustZone (both TrustZone-A and TrustZone-M) does not natively provide this abstraction that enables security-sensitive code to execute in the Normal World with elevated privileges (TrustZone-M does not even feature virtual memory to support user-space processes).
Prior studies~\cite{feng2021scalable,brasser2019sanctuary} introduce enclave-based execution for ARM TrustZone-A.
However, these techniques are not directly applicable to low-end TrustZone-M devices: Penglai~\cite{feng2021scalable} relies on a MMU to enforce isolation, while Sanctuary~\cite{brasser2019sanctuary} requires the deployed devices to be multi-core.
Such hardware features are typically unavailable on low-end edge devices. For the low-end class of devices targeted in this work, several prior efforts have focused on specific security services, including software integrity verification~\cite{eldefrawy2012smart,nunes2019vrased,noorman2013sancus,nunes2024toward}, verifiable function execution~\cite{rattanavipanon2025slapp,nunes2020apex,neto2025pearts}, secure software updates~\cite{de2022casu,de2019pure}, protection of confidential sensor inputs~\cite{nunes2022privacy,de2024sa4p}, and compartmentalization~\cite{ucca}.
While these approaches provide important building blocks, none of them achieves the full set of goals targeted in this work by themselves.
}

\section{Conclusion}

We presented \acro, a system architecture for confidential and verifiable DNN inference on ARM TrustZone-M platforms.
\acro builds on a new execution abstraction, \iobox, which introduces a third execution environment in the Non-Secure World while enforcing strong isolation via Secure-World controls. This design allows inference code to run without expanding the Secure-World TCB, while still protecting sensitive model components and enabling verifiable execution.

\clearpage
\bibliographystyle{ieeetr}
\bibliography{references}

\appendices

\section{Implementation Details}
\label{apdx:impl-details}

We implement \acro on a NUCLEO-L552ZE-Q development board~\cite{st_nucleo_l552ze_q}, which integrates an ARM Cortex-M33 MCU running at 110MHz with 512KB of Flash memory and 256KB of SRAM, where 192KB is used for the Normal World (NS RAM) and the remaining 64KB is reserved for the Secure World (S RAM).

\textbf{Secure World Setup.}
The Secure World runs Trusted Firmware-M (TF-M)~\cite{tfm_project}, a reference Secure Processing Environment for Armv8-M systems. We extend TF-M to implement \acro Secure-World APIs on \dev. Specifically, we use the TF-M Crypto Partition to provide cryptographic services and expose them to the Normal World via NSC veneers. For digital signatures on \dev, we rely on TF-M’s ECDSA APIs (\texttt{PSA\_ALG\_ECDSA}) for signing and verification. At boot, TF-M configures five of the eight SAU regions. Regions 0 and 1 map the Flash and SRAM assigned to the Normal World. Regions 2-4 are reserved for NSC veneers, peripherals, and platform-specific registers, and are left unchanged by our implementation.

When an \iobox instance enters the \texttt{Inactive} state via the \texttt{Create} API, it dynamically reconfigures additional SAU regions to isolate its code and data. For SRAM, Region 1 is disabled, while Regions 6 and 7 are configured to cover all NS RAM except the private-data region, which thus becomes part of Secure state by default. 
For Flash, we reconfigure Region 0 to cover the area preceding the \iobox code and public data, and Region 5 to cover the following region up to Secure Flash. 
Upon transitions to other lifecycle states, the Secure-World APIs restore the SAU configuration to its boot-time defaults.

\textbf{\iobox Setup.}
In our experiments, $\f$ running inside \iobox performs two tasks: (1) acquiring an inference input and (2) executing inference. For (1), we use a simple routine that reads an image from Flash memory to establish a lower-bound timing baseline. A real-world sensor pipeline using an external camera is presented later in Section~\ref{subsec:case-study}.
For (2), we use the CMSIS-NN library~\cite{lai2018cmsis}, which provides optimized DNN kernels with low memory overhead for Arm Cortex-M processors. As our target model, we adopt ResNet~\cite{targ2016resnet}, a lightweight architecture widely used in prior edge-inference work~\cite{11130497,ghosh2023energy,liu2025ed}. Following CMSIS-NN guidelines, we apply fixed-point quantization, resulting in a model size of 76.375KB.
With ResNet, \acro supports arbitrary TSDP partitioning strategies that split the model into $\mPub$ and $\mPriv$. In our evaluation, we follow~\cite{11130497}, designating the final three layers of ResNet as private to reduce the risk of model-stealing attacks. These layers form $\mPriv$ and are provisioned in encrypted form in Flash during \textbf{I1}, while the remaining layers $\mPub$ remain unprotected. 
As such, the full model size is $|\m| = 76.375$\,KB, with $|\mPriv| = 38.625$\,KB and $|\mPub| = 37.75$\,KB.

\section{Discussion, Limitations, and Extensions}\label{apdx:limit}

\textbf{Multiple Inferences per Execution.}
Our case study focuses on producing a proof for a single inference while preserving model confidentiality.
However, the design can be extended to support verifiable execution over multiple sensor inputs within a single invocation.
Rather than invoking \texttt{Execute} repeatedly, $\f$ can be written in a way to accept an additional parameter specifying the number of sensor readings to acquire.
During execution, $\f$ records multiple inputs within the \iobox stack.
After collecting the required number of inputs, $\f$ performs batched inference over the recorded data, producing a vector of outputs.
Since \iobox generates a proof that binds execution to the correct invocation of $\f$, the resulting proof can also attest that the output vector was produced from the specified batch of inputs under a single execution context.
Supporting this extension would additionally require \texttt{Execute} to account for the total number of inferences performed within one invocation when enforcing the authorized usage limit.

\textbf{Interrupt-driven \f.}
To ensure verifiable execution, \acro currently relies on \iobox to disable interrupts during execution of $\f$.
As a result, $\f$ must not depend on interrupt-driven behavior.
While this assumption holds in our case study, it may not apply to applications where $\f$ relies on interrupt-triggered sensor inputs.
One possible extension is to incorporate an interrupt-compatible proof-of-execution mechanism such as~\cite{caulfield2022asap}.
Doing so permits interrupts during execution by enforcing read-only protection over program memory and the interrupt vector table (IVT), while computing a hash of the IVT snapshot before execution begins.
The resulting hash can then be incorporated into $\infToken$ or the generated proof to preserve execution integrity despite interrupt handling.



\textbf{Cryptographic Choice.}
\acro protocol relies on public-key cryptography for authentication and proof generation.
While this overhead is generally acceptable for \modOwner, \modUser, and \vrf, it may be expensive for resource-constrained \dev-s.
In \acro, \dev maintains three keys: $pk_{\modOwner}$, $pk_{\modUser}$, and $sk_{\dev}$.
The first two are used to authenticate requests in \texttt{Authorize} and \texttt{Execute}, respectively, while $sk_{\dev}$ is used to generate \infToken.
To reduce computational overhead on \dev, \acro can replace public-key operations with symmetric-key ones.
Since \modOwner controls provisioning, it may install a shared symmetric key $k_1$ during \texttt{Provision} instead of relying solely on $pk_{\modOwner}$.
Then, during \textbf{I2: Authorization}, \modOwner first authenticates \modUser through standard PKI.
Upon successful authentication, \modUser generates a fresh symmetric session key $k_2$ and securely provides it to \modOwner.
\modOwner then constructs a \texttt{Authorize} request encrypted (via authenticated encryption) using $k_1$, embedding both the authorization policy and $k_2$.
After receiving the request, \dev decrypts it using $k_1$, recovering $k_2$ and using it in place of $(sk_\modUser, pk_\modUser)$ in subsequent interactions.
%
However, replacing $sk_{\dev}$ with a symmetric key introduces additional trust assumptions.
In particular, \vrf would need to share the same secret key with \dev in order to validate proofs, requiring \vrf to be trusted.
This changes the stakeholder model described in Section~\ref{sec:overview}, where \advMod can include \vrf.
Moreover, by design, a symmetric-key option does not support multiple mutually distrusting \vrf-s, since all of them would need access to the same shared secret.
One possible workaround is to designate \modOwner as the sole \vrf.
Under this model, external parties seeking to validate $\infToken$ would query \modOwner, who performs verification on their behalf.



\end{document}